\documentclass[prd,nofootinbib,preprint]{revtex4}
\usepackage[T1]{fontenc}
\usepackage{amsmath,amssymb}
\usepackage{epsfig}
\graphicspath{{Figs/}}
\usepackage{dcolumn}
\usepackage{bm}
\usepackage{graphicx}
\usepackage[usenames,dvipsnames]{color}
\usepackage{slashed}
\usepackage[colorlinks,citecolor=blue]{hyperref}
\usepackage{pdfpages}
\usepackage{float}
\usepackage{adjustbox}
\usepackage[autostyle]{csquotes}
\begin{document}
\title{Octant of $\theta_{23}$, MH,  $0\nu\beta\beta$ decay and  vacuum alignment of  $ A_{4} $ flavour symmetry in an inverse seesaw model}
	
	\author{Maibam Ricky Devi}
	\email{deviricky@gmail.com}
	
	\author{Kalpana Bora}
	\email{kalpana@gauhati.ac.in}
	\affiliation{Department of Physics, Gauhati University, Guwahati-781014, Assam, India}
	
\begin{abstract}
Measurements of disappearance channel of long baseline accelerator based experiments (like NO$\nu$A) are inflicted with the problem of octant degeneracy. In these experiments, the mass hierarchy (MH) sensitivity depends upon the value of CP-violating phase $\delta_{CP}$.  Moreover, MH of light neutrino masses is still not fixed. Also, the flavour structure of fermions is yet not fully understood. We discuss all these issues, in a highly predictive, low-scale inverse seesaw (ISS) model within the framework of $A_4$ flavour symmetry. Recent global analysis has shown a preference for normal hierarchy and higher octant of $\theta_{23}$, and hence we discuss our results with reference to these, and find that  the vacuum alignment of $A_4$ triplet flavon (1,-1,-1) favours these results. Finally, we check if our very precise prediction on $m_{ee}$ and the lightest neutrino mass falls within the range of sensitivities of the neutrinoless double beta decay ($0\nu\beta\beta$) experiments. We note  that when octant of $\theta_{23}$ and MH is fixed by more precise measurements of future experiments, then through our results, it would be possible to precisely identify the favourable vacuum alignment corresponding to the $A_{4}$ triplet field as predicted in our model.
\keywords{Neutrino oscillations; inverse seesaw model; $ A_{4} $ flavour symmetry.}
\end{abstract}
	\maketitle

\section{Introduction}
\label{sec:level1}
Experimental evidences in support of neutrino oscillations during last few decades have led theoretical physicists to construct new models. Recent results from experiments worldwide involving neutrinos have not only ascertained that the neutrinos are massive, with the mixing of flavours while they travel, but also provided precise values of their mixing parameters. Yet these experiments failed to explain the nature of the neutrinos, i.e., if they are Dirac or Majorana, the hierarchy of masses in the three families of neutrino and their absolute values, octant of the atmospheric mixing angle $\theta_{23}$ and the leptonic CP-violating (CPV) phases. It is well known that measurements of disappearance channel of long baseline (LBL) accelerator based experiments (like NO$\nu$A) are inflicted with the problem of octant degeneracy, where the solutions in both octants of $\theta_{23}$ are obtained for both the mass hierarchies.  Moreover, MH of neutrino masses is still not fixed and its sensitivity depends upon the value of CP-violating phase $\delta_{CP}$ (see for example discussions in \cite{Cao:2020ans} , \cite{deSalas:2020pgw}). So, in a way, it can be stated that measurements of octant of $\theta_{23}$, MH and CPV phase are dependent on each other (or we can say that they are entangled).\\
\\
It is well known that the models based on type-I and type-II seesaw mechanism \cite{Penedo:2017knr} are very unlikely to get tested in experiments due to the presence of very heavy right-handed neutrinos whose mass scale can go up to ($\sim 10^{10}-10^{14} $) GeV. To circumvent this problem, one can construct low-scale seesaw models with Majorana neutrinos of a relatively lower mass scale of few TeV. Such models are much more accessible to get tested in ongoing neutrino experiments to find new physics, and this is the motivation behind considering low-scale seesaw models. Also, flavour structure of fundamental particles still remains unexplained. Hence in this work, we choose to consider an ISS model with $A_4$ flavour symmetry \cite{Ishimori:2012zz} .Many flavour symmetries have been used by researchers for the purpose, however, no information is available so far about the scale of flavour  symmetry breaking or the direction in which flavon fields acquire VEV.\\
\\
To obtain some insight about VEV alignment of flavon field of  $A_{4}$ flavour symmetry, we use the information available on neutrino oscillation parameters from various experiments. To be specific, we use the information on octant-MH degeneracy present in the measurements of disappearance channels of NO$\nu$A \cite{Cao:2020ans} , T2K \cite{Abe:2011sj} etc, and the fact that MH among neutrino masses is yet not fixed. Though disappearance channel suffers from this degeneracy, appearance channel does not. We would like to mention here that this work is an extension of our previous work \cite{Devi:2021ujp} . The Lagrangian is constructed up to dimension 6 with mass term $ \mu_{s} $, as we need to make sure that $ \mu_{s} $ has a very low energy scale compared to the other mass terms. Some extra symmetries such as $ Z_{4} $, $ Z_{5} $ and $ U(1)_{X} $ are integrated along with the $ A_{4} $ symmetry to make the model feasible where  $ U(1)_{X} $  is a global symmetry. We obtain the light neutrino mass matrix from the Lagrangian, which when compared to the phenomenological one, gives us a set of equations for the triplet and singlet flavon fields. These equations are then solved by using the known values of mixing angles and the neutrino mass differences as inputs, to find the unknown parameters such as the  $ m_{lightest}$ and CP-violating phases $ \delta_{CP} $(Dirac phase), $ \alpha $ and $ \beta$ (Majorana phases). This makes our model highly testable with a heavy Majorana neutrino of $ \approx $1 TeV. We constrain the values of the known parameters within the 3$ \sigma$ ranges of their latest global best fit data which are summarised in Table (\ref{tab:data1}). The new values of unknown oscillation parameters predicted here can be used in pinpointing the favoured vacuum alignment of triplet flavon field of $A_4$ symmetry. Next, these are used to study  implication on effective neutrino mass $m_{ee}$ measured in $0\nu\beta\beta $ experiments,  and correlations between lightest neutrino mass $m_{lightest}$ and $m_{ee}$ are obtained.  All these interesting results along with low scale of seesaw mechanism are testable in future experiments.\\
\\
Some earlier works on $ A_{4} $ flavour symmetry based neutrino models can be found in \cite{Dinh:2016tu,Chen:2012st, Kalita:2015jaa, Sarma:2018bgf} . $A_{4}$ based inverse and linear seesaw models at low scale can be found in \cite{Hirsch:2009mx, Sruthilaya:2017mzt} , where in \cite{Hirsch:2009mx}  a linear as well as an inverse seesaw models incorporating $SU(2)_L $, $Z_{3}$ and $ A_{4} $ symmetries was discussed. Similarly, 
in \cite{Sruthilaya:2017mzt} , a discussion on linear seesaw model with $ A_{4} $ flavour symmetry, $   Z_{4}$, $ Z_{3} $ discrete symmetries and a $U(1)_{X}$ global symmetry is presented. In  \cite{Borah:2017dm} , $ A_{4} $ symmetry based type I as well as inverse seesaw model was considered and in \cite{Sahu:2020tqe} an $ A_{4} $ LRSM was discussed. To study other flavour symmetry based neutrino models one can also refer to the works in \cite{Verma:2018lro, Girardi:2015rwa, Petcov:2018snn, Petcov:2018mvm, Borah:2018nvu, Sethi:2019bfu, Cai:2018upp} . Since we are interested to link resolving of octant-MH degeneracy with VEV alignment of flavon field, it is relevant to understand the importance of the former issue, for the sake of pedagogy, and for that purpose, we take a note of some earlier works to resolve this.  Several papers have discussed the current status of light neutrino oscillation parameters, and methods have been suggested earlier to resolve octant-MH degeneracy \cite{Cao:2020ans, Rahaman:2021cgc, Rout:2020cxi, Yasuda:2020cff, Ghosh:2019sfi, Haba:2018klh, Verma:2018gwi, Bharti:2018eyj, Fogli:1996pv, Barger:2001yr, Smirnov:2018jzl, Bora:2014zwa} . A recent discussion on 2$ \sigma $ tension in T2K and NO$ \nu $A can be found  in \cite{deSalas:2020pgw, Denton:2020uda, Kelly:2020fkv, Esteban:2020itz, alex_himmel_2020_3959581, Nizam:2018got} . \\

\begin{table}
\caption{Updated values of neutrino oscillation parameters in 3$ \sigma $ values taken from \cite{deSalas:2020pgw}.}
{\begin{tabular*}{\columnwidth}{@{\extracolsep{\fill}}lll@{}}
\hline
\multicolumn{1}{@{}l}{Neutrino Oscillation Parameters} & for normal hierarchy (NH) & for inverted hierarchy (IH)\\
\hline
\\
$ \Delta m_{21}^{2}/(10^{-5} eV^{2})$ & $6.94-8.14$ & $6.94-8.14 $ \\
$ |\Delta m_{31}^{2}|/(10^{-3} eV^{2})$ & $2.47-2.63$ & $2.37-2.53$ \\
$ \theta_{12}/^{\circ} $ &  $31.4-37.4$ & $31.4-37.4 $ \\
$ \theta_{23}/^{\circ}  $ & $ 41.20-51.33$ &  $41.16-51.25 $ \\
$\theta_{13}/^{\circ}  $ & $8.13-8.92$ & $8.17-8.96 $ \\
$\delta_{CP} /^{\circ} $ & $128-359$ & $200-353 $\\
\\
\hline
\end{tabular*}
\label{tab:data1}}
\end{table}

It is worth mentioning however, that in a very recent global analysis \cite{deSalas:2020pgw}, they have presented some preferences for higher octant of atmospheric mixing angle and normal mass hierarchy, and we have discussed our results in context of this analysis too and  found that the VEV (1,-1,-1)/(-1,1,1) of the $A_4$ triplet flavon field is preferred. Such information on pinpointing VEV of flavour symmetry has not been presented earlier. This novel information is then used to predict very precise values of $m_{lightest}$ and $m_{ee}$ (of $0\nu\beta\beta$ decay), which can be tested in future. We have done our computation  up to an accuracy of $10^{-5}$, which indicates the precision level of the computation.\\

The paper has been organised as follows. A brief discussion on octant-MH degeneracy and the current status of values of neutrino oscillation parameter values is presented in section \ref{sec:level2}. In section \ref{sec:3}, we present the implication of neutrino oscillation parameters on neutrinoless double beta decay. We discuss our inverse seesaw model with $ A_4 $ symmetry in section \ref{sec:level3} and in section \ref{sec:level4}, the numerical analysis and results are presented.  A discussion on our results is summarised in section \ref{sec:level5} and then we conclude in section \ref{sec:level6}. 


\section{Octant-MH degeneracy and current status of neutrino oscillation parameters}

\label{sec:level2}
The Super-Kamiokande experiment (SK) \cite{Super-Kamiokande:1998kpq} and Sudbury Neutrino Observatory (SNO) \cite{SNO:2001kpb, SNO:2002tuh} after discovering neutrino oscillation phenomena, provided us strong evidence that  neutrinos are massive particles whose mass can not be explained within Standard Model, and that there is a mixing of flavours while they travel. The probability in the disappearance channel of the long baseline experiments can be expressed as \cite{Bora:2014zwa} , \\
\begin{equation}
P_{\mu \mu} = 1-sin^{2}2\theta_{23}sin^{2}\dfrac{1.27 \Delta m^{2}_{31} L}{E}  +4 sin^{2}\theta_{13}sin^{2}\theta_{23}cos{2\theta_{23}}sin^{2}\dfrac{1.27\Delta m^{2}_{31}L}{E}
\label{Eqn 1}
\end{equation}
where $ \Delta m^2_{ij} = m_i^2-m_j^2$, $\Delta_{ij}=\dfrac{\Delta m^2_{ij}L}{4E}$ and $\dfrac{\Delta^{2}_{21}}{\Delta^{2}_{31}} $ is assumed to be  small. The measurements of disappearance channel of long baseline experiments cannot differentiate the octant of the atmospheric mixing angle, as it is clear from Eq. (\ref{Eqn 1}) that the probability is sensitive to $ sin^{2}2\theta_{23} $ and hence $ P(\theta_{23}) =P(\pi/2 -\theta_{23}) $. This is called octant degeneracy as one cannot differentiate if  $ \theta_{23} > \pi/4$ or  $ \theta_{23}<\pi/4 $  \cite{Fogli:1996pv, Barger:2001yr} (or in other words it cannot differentiate if  $ sin^{2}\theta_{23} > 0.5$ or  $ sin^{2}\theta_{23}<0.5 $) \cite{deSalas:2020pgw}  from experimental measurements. The LBL experiments are also quite sensitive to measurement of CP-violating phase $ \delta_{CP}$.  Recent measurements have shown that $ \delta_{CP} \approx 0.8 \pi $ from NO$ \nu $A analysis while it disfavours the region $ \delta_{CP} \approx 1.5 \pi $ which coincides with the T2K best fit values \cite{deSalas:2020pgw} . Hence, there is ambiguity in CPV phase measurements at T2K and NO$\nu$A. The atmospheric neutrino results from Super-Kamiokande \cite{Super-Kamiokande:2017yvm} and Deep Core (Ice Cube) experiments \cite{IceCube:2017lak, IceCube:2019dqi} when are combined with the data of the long baseline experiments \cite{Denton:2020uda, Kelly:2020fkv, Esteban:2020itz} , then substantially more preference for normal hierarchy is observed. However, as the measurements of $ \delta_{CP} $ in T2K and NO$ \nu $A are inflicted with discrepancy as discussed above, no clearcut preference for a particular MH is seen in their data.  In a recent work \cite{Cao:2020ans} ,  it was shown that the data of T2K, NO$ \nu $A and JUNO can be combined to enhance the measurement of $ \delta_{CP} $ and resolve MH and octant degeneracy problem too. \\
\\
Thus it is seen that the neutrino oscillation experiments face the problem of parameter  degeneracies and measurement of one parameter depends on that of another, and it is still a challenge before theorists as well as experimentalists alike.  Many proposals have been given to resolve these ambiguities  \cite{Bharti:2018eyj, Bora:2014zwa} .  Suppose in future, these parameter degeneracies are resolved, then, we propose in this work, that  we can identify the vacuum alignment of the triplet flavon field. It may be noted that at present, no information about flavour symmetries is available, and hence findings of this work command importance. 
\section{Implication of neutrino oscillation parameters on neutrinoless double beta decay}
\label{sec:3}
The search for $ 0\nu\beta\beta $ decay is very intriguing as its detection will confirm the Majorana nature of neutrinos. In this process, neutrinos are not emitted due to the immediate reabsorption of the emitted neutrinos. These neutrinos thus have an effective mass that can be expressed as \cite{Rodejohann:2011mu, Bilenky:2014uka} ,\\
\begin{equation}
 m^{\nu}_{ee} = |cos\theta^{2}_{12}
cos\theta^{2}_{13} m_{1} + sin\theta^{2}_{12} cos\theta^{2}_{13}
m_{2} e^{2i\alpha} + sin\theta^{2}_{13} m_{3} e^{2i\beta}| \textrm{.}
\label{effective 0bb mass}
\end{equation}\\
Thus $  m^{\nu}_{ee} $ depends exclusively on the neutrino oscillation parameters that can be obtained from our model as well as from the 3$ \sigma $ values of global neutrino oscillation data. The half-life of the isotope $ T^{0\nu}_{1/2}(\mathcal{N}) $ which is involved in the $ 0 \nu \beta \beta $ decay is constrained by its decay amplitude  \cite{deSalas:2020pgw, DellOro:2016tmg} . This lifetime of isotopes can constrain the bounds on the $ m^{\nu}_{ee} $ from the $ 0\nu \beta \beta $ events as \cite{deSalas:2020pgw}
\begin{equation}
T^{0\nu}_{1/2}(\mathcal{N})  =\dfrac{m^{2}_{e}}{G^{\mathcal{N}}_{0\nu}| \mathcal{M}^{\mathcal{N}}_{0\nu}|^{2}m^{2}_{\beta \beta}}
\end{equation} \\
where the electron mass is denoted by $ m_{e} $, $G^{\mathcal{N}}_{0\nu}$ and $\mathcal{M}^{\mathcal{N}}_{0\nu}$ represents the factor involving phase space and the nuclear matrix element that depends on the isotope used in the experiment respectively. The strongest bounds on half-life $ T^{0\nu}_{1/2}(\mathcal{N}) $ set by GERDA \cite{GERDA:2019ivs} , CUORE \cite{CUORE:2019yfd} and KamLAND-Zen \cite{KamLAND-Zen:2016pfg} are $ T^{0\nu}_{1/2}> 9 \times 10^{25}$ years for $ ^{76} $Ge, $ T^{0\nu}_{1/2}> 3.2 \times 10^{25}$ years for $ ^{130} $Te and $ T^{0\nu}_{1/2}> 1.07 \times 10^{26}$ years for $ ^{136} $Xe respectively at $ 90\% $ confidence level. 
Different $ 0\nu\beta\beta $ experiments use different experimental techniques such as external trackers, bolometers, semiconductor detectors to probe this rare decay.\\
The current upper limits of $ m_{\beta \beta} $ are $m_{0\nu\beta\beta} < $  104-228 meV by GERDA \cite{GERDA:2019ivs} ,  $m_{0\nu\beta\beta} < $  75-350 meV by CUORE \cite{CUORE:2019yfd} experiment, and $m_{0\nu\beta\beta} < $  61-165 meV by KamLAND-Zen  experiment \cite{KamLAND-Zen:2016pfg} . 
Some currently operating experiments are CANDLES III, COBRA, MAJORANA Demonstrator, CUORE, NEXT WHITE and KamLand - Zen 800 \cite{Calibbi:2017uvl} . For these ongoing as well experiments under construction such as SuperNEMO, LEGEND 200, Amore I and Amore II, NEXT etc., the future sensitivity of $ m_{\beta \beta} $ is expected to reach up to $ 5\times 10^{-3} $ eV. In later sections, we compute  $m_{0\nu\beta\beta}$ allowed in our model and find that it agrees well with its current upper limits of some of these experiments.


\section{ An $ A_{4}\times U(1)_{X} \times Z_{5} \times Z_{4} $ inverse seesaw model }
\label{sec:level3}
To develop our model based on seesaw mechanism  \cite{PhysRevLett.56.561, PhysRevD.34.1642} , we use cyclic groups $ Z_{4} $ and $ Z_{5} $ along with $A_{4}$ group and $ U(1)_{X} $ global symmetry. Our model includes an extra family of three singlet sterile neutrinos, $  \mathcal{S}$ apart from the SU(2) singlet neutrinos, $  \mathcal{N}$. Thus we can write the neutrino mass matrix obtained from the Lagrangian of the seesaw model in the basis ($ \nu^{c}_{L}, \mathcal{N}, \mathcal{S} $) as \cite{Wyler:1982dd} :\\
\begin{equation}
 \left( M_{\nu} \right)_{iss}=\begin{bmatrix}
0 & m_{D} & 0\\
m^{T}_{D} & 0 & M\\
0 & M^{T} & \mu_{s}
\end{bmatrix} \textrm{.}  
\label{ISS: ISS mass matrix}
\end{equation}\\
To implement the inverse seesaw mechanism, the necessary condition $m_{D}, M \gg \mu_{s}$ should be satisfied. This condition is required to generate the light neutrino mass matrix in $ \mathcal{O}(eV) $ which is doubly suppressed by M as, 
\begin{equation}
m_{\nu}= m_{D}(M^{T})^{-1}\mu_{s} M^{-1}m^{T}_{D}\textrm{ .}
\label{ISS:neutrino_matrix_equation}
\end{equation}
\\
Here the $\mu_{s}$ term of the low-scale seesaw mechanism breaks down the lepton number conservation\cite {tHooft:1980xss} .  We present the particle content of the various fields considered in the model in Table (\ref{tab:vevA4ISS}) where $\Phi_s$, $\Phi_t$ are triplet scalar fields and $ \eta $, $\xi$, $\rho$, $\kappa$, $\kappa^{\prime}$, $\kappa^{\prime \prime}$ are singlet scalar fields under $ A_{4} $ group transformation.  $\mathcal{L}$ is the SM LH leptonic doublet and $\mathcal{H} $ is the SM Higgs doublet.

\vspace{0.1 in}
\begin{table}[h]
\caption{Transformation of the fields in our model under $ A_{4}\times Z_{4} \times Z_{5} \times U(1)_{X} $ symmetry for neutrino mass model realising inverse seesaw mechanism.}
{\begin{adjustbox}{width=10cm}
\begin{tabular}{|c|cc|cc|cc|ccc|cccccc|}
\hline
 & $\mathcal{L}$ & $\mathcal{H} $ & $\Phi_t$ & $\Phi_s$ &  $\mathcal{N}$ & $ \mathcal{S}$ & $ \mathcal{e}_{r} $ & $ \mathcal{\mu}_{r} $ & $ \mathcal{\tau}_{r} $ &  $\eta$ & $\xi$ & $\rho$ & $\kappa$ & $\kappa^{\prime}$ & $\kappa^{\prime \prime}$\\
\hline
$A_4$ & 3 & 1 &3 & 3 & 3 & 3 & 1 & $ 1^{\prime\prime} $ & $ 1^{\prime} $ &  1 & $1'$ & $1''$ & 1 & 1 & 1 \\
\hline 
$U(1)_{X}$ & -1 & 0 & 0 & -1 & -1 &  1 & -1 & -1 & -1 &  -1 & -1 & -1 & 0 & -4 & -3 \\
\hline
$Z_{5}$ & 1  & 1 & $ \omega $ & 1  & $ \omega^{2} $ & 1 & $ \omega$ & $ \omega $ & $ \omega $ & 1 & 1 & 1 & $ \omega^{2} $ & 1 & 1 \\
\hline
$Z_4$ & 1 & 1 & i & -i  & i & 1 & i & i & i & -i & -i & -i & i & i & 1 \\
\hline

\end{tabular}
\end{adjustbox}
\label{tab:vevA4ISS}}
\end{table}

If we now apply the $ A_{4} $ product rules on above fields, we obtain the Lagrangian for the charged leptons as:\\
\begin{equation} 
\mathcal{L}_{c.l.} \supset  \dfrac{y_{1}}{\Lambda}(\bar{\mathcal{L}}\Phi^{\dagger}_{t})\mathcal{H} \mathcal{e}_{r}+\dfrac{y_{2}}{\Lambda}(\bar{\mathcal{L}}\Phi^{\dagger}_{t})^{\prime}\mathcal{H} \mathcal{\mu}_{r}+\dfrac{y_{3}}{\Lambda}(\bar{\mathcal{L}}\Phi^{\dagger}_{t})^{\prime\prime}\mathcal{H} \mathcal{\tau}_{r}\textrm{ .}
\label{ISS:revised charged Lagrangian modified}
\end{equation}\\
Eq. (\ref{ISS:revised charged Lagrangian modified}) can be expressed in matrix form as:\\
\begin{equation}
 M_{c.l.}=\dfrac{\upsilon^{\dagger}_{t}\upsilon_{h}}{\Lambda}\begin{bmatrix}
y_{1} & 0 & 0\\
0 & y_{2} & 0\\
0 & 0 & y_{3}
\end{bmatrix} \textrm{,}
\label{Charged lepton mass matrix}
\end{equation}\\
Here we have taken the VEVs of the standard model Higgs as $ \langle h \rangle= \upsilon_{h} $ and $ \langle \Phi_{t} \rangle= \upsilon_{t} $ respectively, $ y_{1} $, $ y_{2}$ and $ y_{3} $ represent the coupling constants  and $ \Lambda $ denotes the usual cut-off scale of the theory. We can proceed to write the Lagrangian of the neutrino sector after the application of $ A_{4} $ product rules: $1'\times 1'=1''$, $1'\times 1''=1, 1''\times 1''= 1'$ and $3\times 3=1+1'+1''+3_A+3_S$ \cite{Altarelli:2010gt} . This can be written as\\
\begin{equation} 
\begin{aligned}
\mathcal{L}_{\rm y} \supset  y_d \frac{\bar{\mathcal{L}} \tilde{\mathcal{H}} \mathcal{N} \kappa^{\dagger}}{\Lambda} + y_m \mathcal{N} \mathcal{S} \kappa^{\dagger} + y_{\mu} \mathcal{S} \mathcal{S} [\frac{\kappa^{\prime}\kappa^{\prime\prime^{\dagger}}(\Phi_s + \eta + \xi + \rho) }{\Lambda^2}  ] + h.c. \textrm{,}
\end{aligned}
\label{ISS:revised Lagrangian}
\end{equation}
\\
where  $  y_d $, $ y_m $, $ y_\mu $ are the coupling constants corresponding to the mass terms $m_{D}$, M and $ \mu_{s} $ in Eq. (\ref{ISS:revised Lagrangian}) respectively.  The VEVs obtained by the scalar fields apart from $\langle h \rangle  $ and $\langle \Phi_{t} \rangle $ after spontaneous symmetry breaking are $\langle \Phi_s \rangle= \upsilon_{s}(\Phi_{a}, \Phi_{b}, \Phi_{c}) $, $ \langle \eta \rangle = \upsilon_{\eta}$, $ \langle \xi \rangle = \upsilon_{\xi}$, $ \langle \rho \rangle = \upsilon_{\rho}$, $ \langle \kappa \rangle = \upsilon_{\kappa}$, $ \langle \kappa^{\prime} \rangle = \upsilon_{\kappa^{\prime}}$ and $ \langle \kappa^{\prime \prime} \rangle = \upsilon_{\kappa^{\prime \prime}}$. The block matrices M, $ m_{D} $ and $ \mu_{s} $ can be expressed in matrix form as:
\\
\begin{equation}
 m_{D}=\dfrac{y_{d}\upsilon_{h}\upsilon^{\dagger}_{\kappa}}{\Lambda}\begin{bmatrix}
1 & 0 & 0\\
0 & 0 & 1\\
0 & 1 & 0
\end{bmatrix} \textrm{,}
\label{ISS: MD}
\end{equation}\\
 \begin{equation}
  M=y_{m}\upsilon^{\dagger}_{\kappa}\begin{bmatrix}
1 & 0 & 0\\
0 & 0 & 1\\
0 & 1 & 0
\end{bmatrix}  \textrm{,}
\label{ISS:M}
\end{equation}
and,
\begin{equation}
  \mu_{s}=\dfrac{y_{\mu}\upsilon_{\kappa^{\prime}}\upsilon^{\dagger}_{\kappa^{\prime\prime}}}{\Lambda^{2}}\left(
\begin{array}{ccc}
 \upsilon_{\eta }+2 \upsilon_s \phi _a & \upsilon_{\xi }-\upsilon_s \phi _c & \upsilon_{\rho }-\upsilon_s \phi _b \\
 \upsilon_{\xi }-\upsilon_s \phi _c & \upsilon_{\rho }+2 \upsilon_s \phi _b & \upsilon_{\eta }-\upsilon_s \phi _a \\
 \upsilon_{\rho }-\upsilon_s \phi _b & \upsilon_{\eta }-\upsilon_s \phi _a & \upsilon_{\xi }+2 \upsilon_s \phi _c \\
\end{array}
\right) \textrm{.}
\label{ISS:Mu}
\end{equation}\\
Substituting these matrices in the Eq. (\ref{ISS:neutrino_matrix_equation}) we obtain,
\begin{equation}
 \Rightarrow m_{\nu}=F\left(
\begin{array}{ccc}
 \upsilon_{\eta }+2 \upsilon_s \phi _a & \upsilon_{\xi }-\upsilon_s \phi _c & \upsilon_{\rho }-\upsilon_s \phi _b \\
 \upsilon_{\xi }-\upsilon_s \phi _c & \upsilon_{\rho }+2 \upsilon_s \phi _b & \upsilon_{\eta }-\upsilon_s \phi _a \\
 \upsilon_{\rho }-\upsilon_s \phi _b & \upsilon_{\eta }-\upsilon_s \phi _a & \upsilon_{\xi }+2 \upsilon_s \phi _c \\
\end{array}
\right)\textrm{,}
\label{ISS:final matrix with vev}
\end{equation}
where, $ F= \dfrac{y^{2}_{d}y_{\mu}}{y^{2}_{m}}(\dfrac{\upsilon^{2}_{h}\upsilon_{\kappa^{\prime}}\upsilon^{\dagger}_{\kappa^{\prime \prime}}}{\Lambda^{4}})$ is a constant of the neutrino mass matrix that depends on the scale of the three mass terms of the Lagrangian and various coupling constants of the theory, as given in Eq. (\ref{ISS:revised Lagrangian}).

\section{Numerical method and results}
\label{sec:level4}

\begin{figure*}
  \centering
\includegraphics[width=1.0\textwidth]{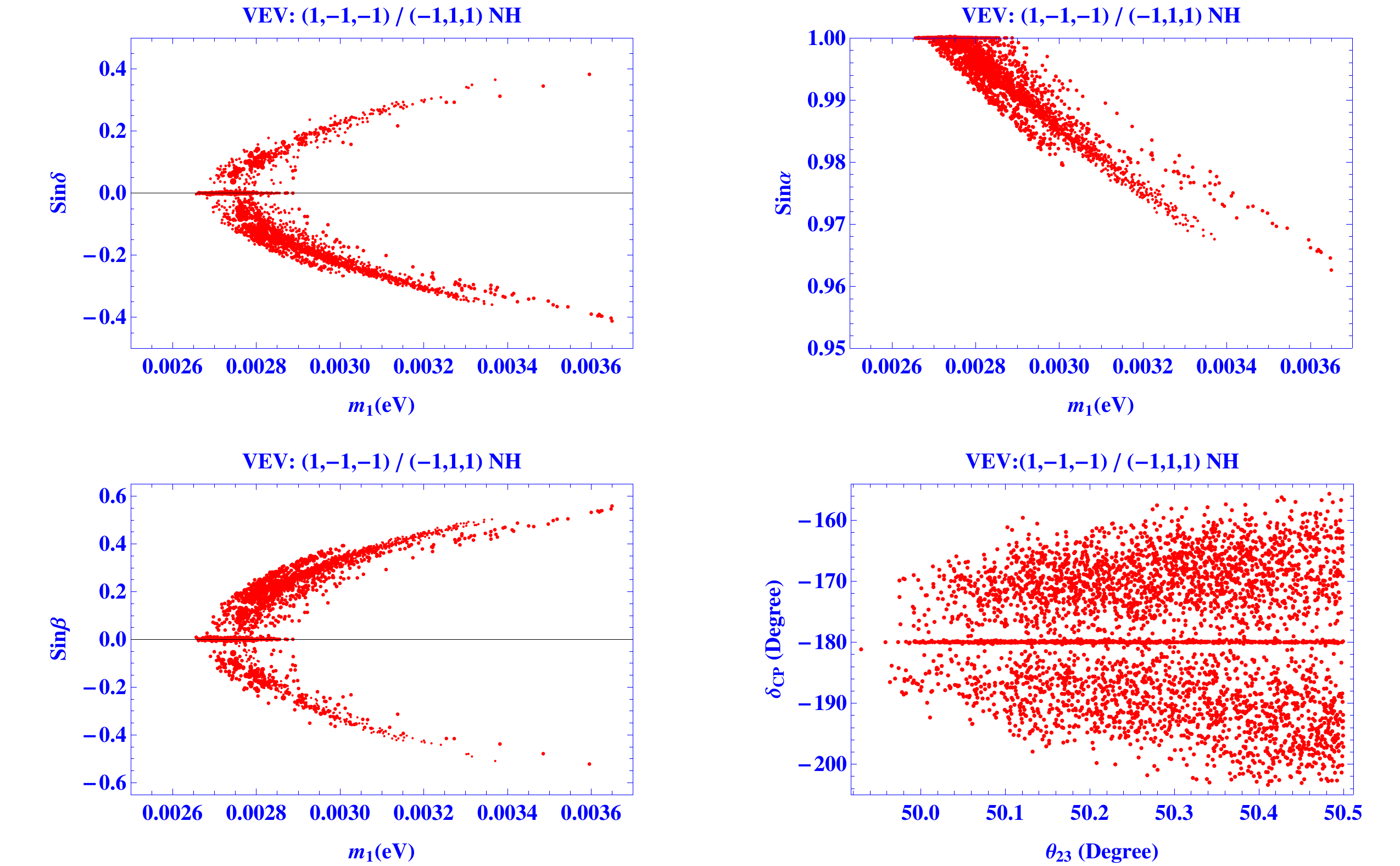} 
 \\
 \caption{The first three plots shows the correlation between the lightest neutrino mass, $ m_{1} $ and the sinusoidal of CPV phases $ \delta_{CP} $, $ \alpha $ and $ \beta $ respectively and the fourth plot shows the correlation between  the atmospheric mixing angle, $ \theta_{23} $ and $ \delta_{CP} $  phase. All of them are plotted for VEV alignment $(1,-1,-1)$ or $ (-1,1,1) $ with normal hierarchy.
 }
 \label{fig13:LSS_NH1-1-1}
\end{figure*}

\begin{figure*}
  \centering
\includegraphics[width=1.0\textwidth]{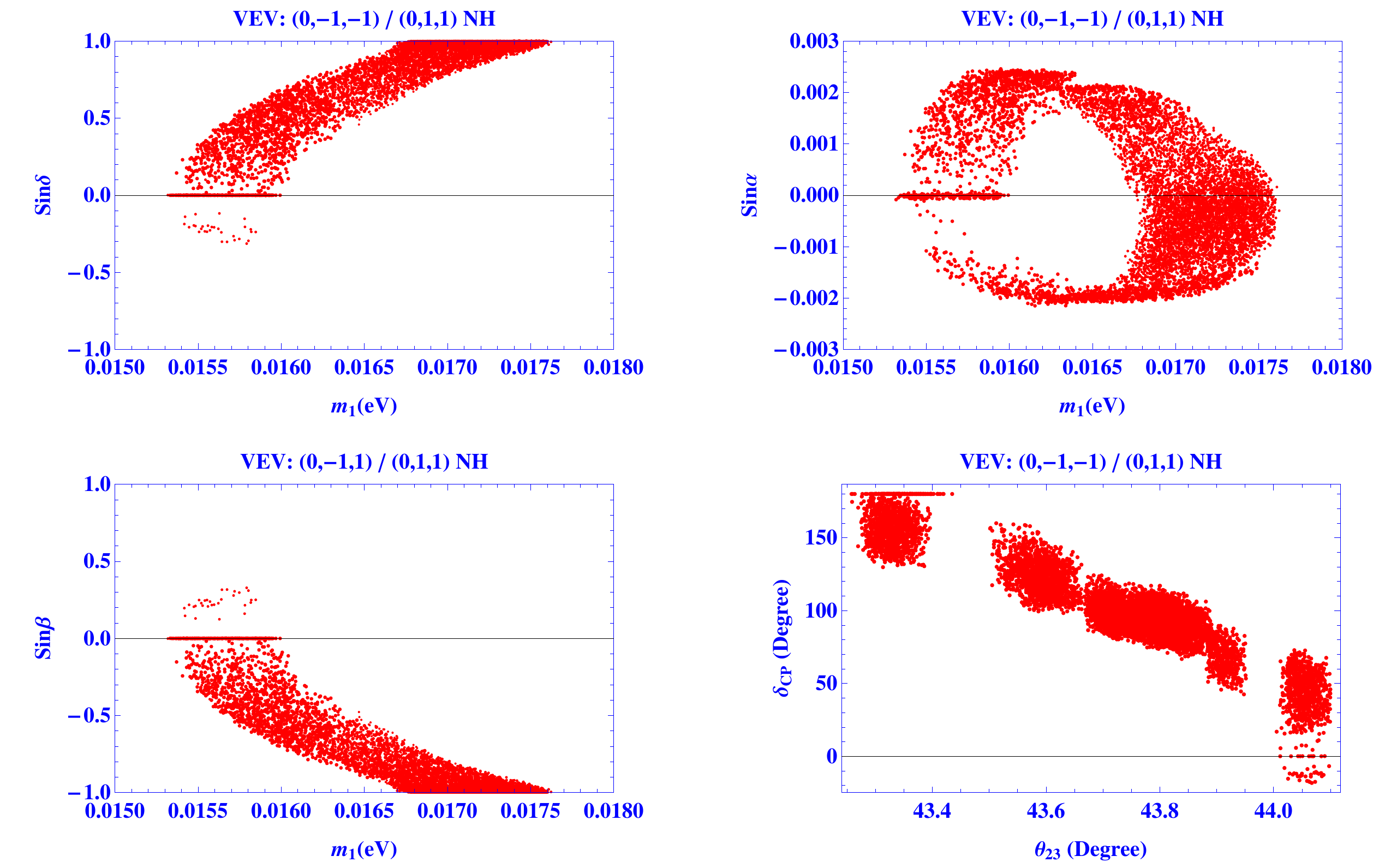} 
 \\
 \caption{
The first three plots shows the correlation between the lightest neutrino mass, $ m_{1} $ and the sinusoidal of CPV phases $ \delta_{CP} $, $ \alpha $ and $ \beta $ respectively and the fourth plot shows the correlation between  the atmospheric mixing angle, $ \theta_{23} $ and $ \delta_{CP} $  phase. All of them are plotted for VEV alignment $(0,-1,-1)$ or $ (0,1,1) $ with normal hierarchy.
}
 \label{fig15:LSS_NH0-1-1}
\end{figure*}

\begin{figure*}
  \centering
\includegraphics[width=1.0\textwidth]{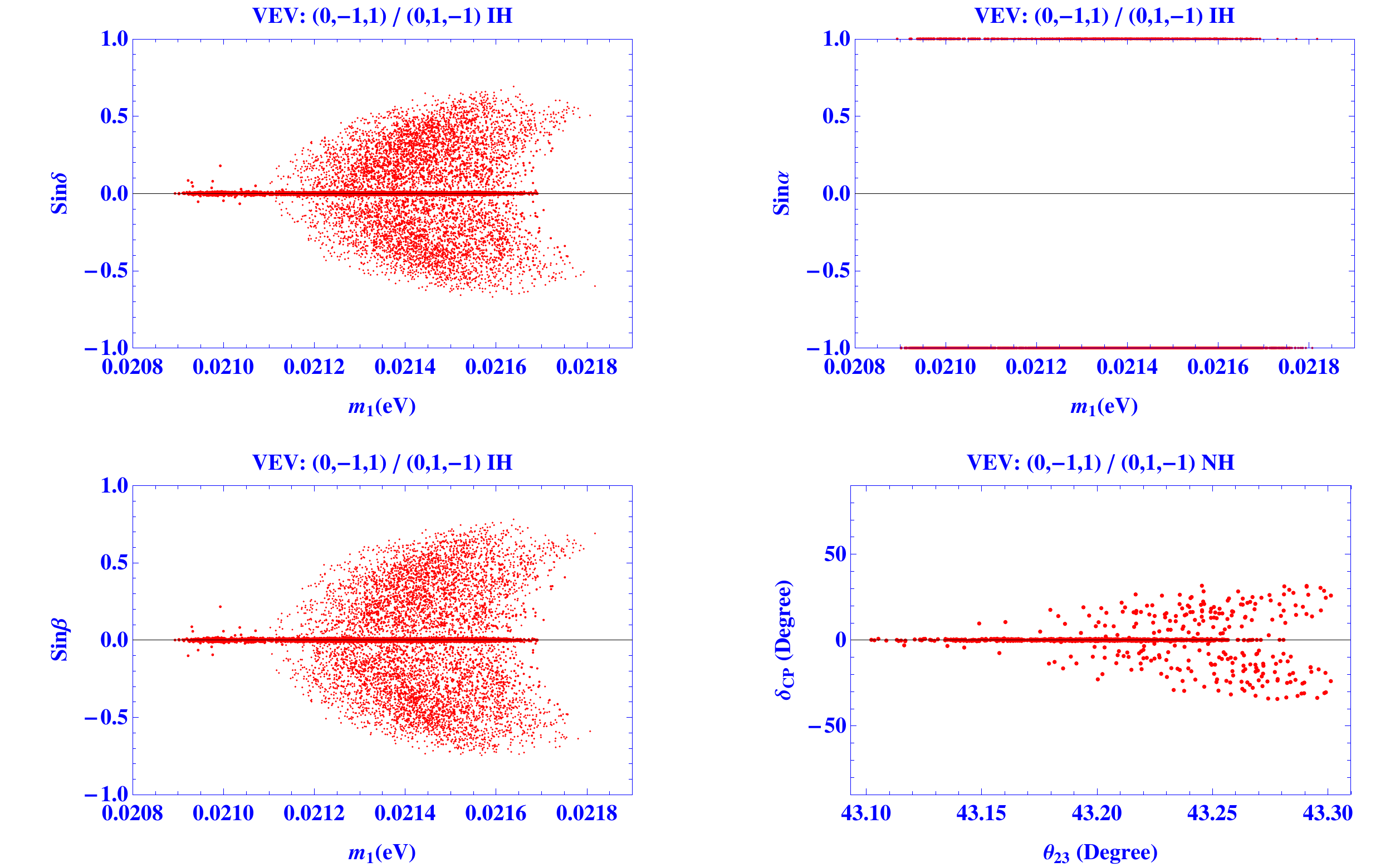} 
 \\

 \caption{The first three plots shows the correlation between the lightest neutrino mass, $ m_{3} $ and the sinusoidal of CPV phases $ \delta_{CP} $, $ \alpha $ and $ \beta $ respectively and the fourth plot shows the correlation between  the atmospheric mixing angle, $ \theta_{23} $ and $ \delta_{CP} $  phase. All of them are plotted for VEV alignment $(0,-1,1)$ or $ (0,1,-1) $ with inverted hierarchy.
}
 \label{fig17:LSS_IH0-11}
\end{figure*}

\begin{figure*}
  \centering
\includegraphics[scale=0.5]{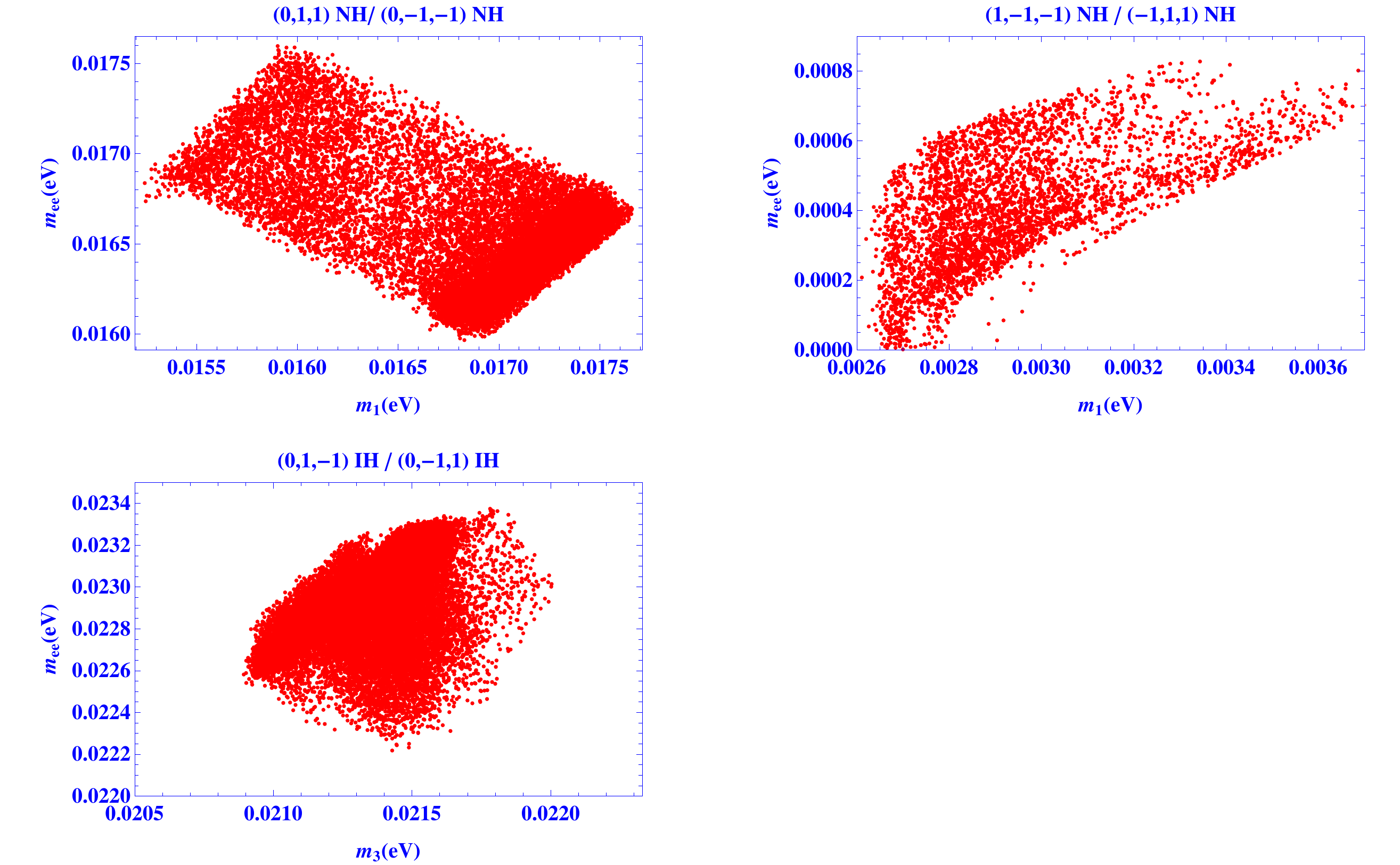}
\caption{ Plot showing the correlation between $ m_{lightest} $ and $ m_{ee} $ of $ 0\nu \beta \beta $ decay for all the allowed vacuum alignments of $ \Phi_{s} $.}
\label{fig:LSS 0vbb}
\end{figure*}

To obtain the undetermined neutrino oscillation parameters such as the lightest mass of the neutrino, CP-violating phases (both Dirac ($ \delta_{CP} $) and Majorana ($ \alpha $, $ \beta $) phases), we first compare the matrix in Eq. (\ref{ISS:final matrix with vev}) with the matrix of light neutrino mass obtained after parametrisation by the PMNS mixing matrix,  $ U_{P} $ i.e.,
\begin{equation}
m_{\nu}=U_{P}. m_{\nu_{diag}}. U^T_{P} \textrm{ .}
\label{pmns0}
\end{equation}
The $ U_{P} $ matrix can be parametrised as:
\begin{equation}
U_{P}=\left(
\begin{array}{ccc}
 1 & 0 & 0 \\
 0 & c_{23} & s_{23} \\
 0 & -s_{23} & c_{23} \\
\end{array}
\right).\left(
\begin{array}{ccc}
 c_{13} & 0 & e^{-i \delta_{CP} } s_{13} \\
 0 & 1 & 0 \\
 -e^{i \delta_{CP} } s_{13} & 0 & c_{13} \\
\end{array}
\right).\left(
\begin{array}{ccc}
 c_{12} & s_{12} & 0 \\
 -s_{12} & c_{12} & 0 \\
 0 & 0 & 1 \\
\end{array}
\right).\left(
\begin{array}{ccc}
 1 & 0 & 0 \\
 0 & e^{i \alpha } & 0 \\
 0 & 0 & e^{i (\beta +\delta_{CP} )} \\
\end{array}
\right) \textrm{ ,}
\label{matrixPMNS}
\end{equation}\\
The mixing angles in Eq. (\ref{matrixPMNS}) can be denoted as $c_{ij}$=$\cos{\theta_{ij}}$,  $s_{ij}$=$\sin{\theta_{ij}}$ and $m_{\nu_{diag}}$$=$$diag(m_1, m_2, m_3)$ which can be reduced in the form  $m^{diag}_{\nu} = diag(m_1, \sqrt{m^2_1+\Delta m_{21}^2}, \sqrt{m_1^2+\Delta m_{31}^2})$ and
 $m^{diag}_{\nu} = diag(\sqrt{m_3^2+\Delta m_{23}^2-\Delta m_{21}^2}, \sqrt{m_3^2+\Delta m_{23}^2}, m_3)$ for normal and inverted mass hierarchy respectively.\\
\\
On comparison of these matrices, a set of six flavon equations are obtained that depend purely on the neutrino oscillation parameters (both known and unknown). Keeping the mixing angles and mass-squared differences as known neutrino oscillation parameters, we solve these equations simultaneously to obtain the undetermined parameters as mentioned previously in this section. We use data for $ \theta_{12} $, $ \theta_{13} $, $ \theta_{23} $ and $ \Delta m_{21}^2 $, $ \Delta m_{31}^2$ and $\Delta m_{23}^2$ from the recent global fit data obtained from various neutrino oscillation experiments  \cite{deSalas:2020pgw} . We solve these equations for different vacuum alignments of $ \upsilon_{s}(\Phi_{a}, \Phi_{b}, \Phi_{c}) $, and have followed numerical analysis as in earlier works  \cite{Chen:2012st, Kalita:2015jaa} and \cite{Sarma:2018bgf} . Our results have been summarised in Table (\ref{tab:para range}) and in Figs (\ref{fig13:LSS_NH1-1-1}-\ref{fig:LSS 0vbb}), we present the correlation between the determined and undetermined neutrino parameters using scattered plots.  Latest global fit results from Ref.\cite{deSalas:2020pgw}  have been summarised in Table (\ref{tab:3}), for comparison with our results. An earlier summary of $0\nu\beta\beta$ decay can also be found in Ref.\cite{DellOro:2014ysa} .

\begin{table}[] \centering
\caption{Summary of allowed and disallowed vacuum alignments of the flavon $ \Phi_{s} $ for neutrino oscillation parameters constrained within their latest $ 3\sigma $ range.}
{\begin{small}
\begin{tabular}{|c|c|}
\hline
\textbf{Allowed VEV in NH} & \textbf{Allowed VEV in IH}  \\ \hline
(0,1,1), (1,-1,-1), (-1,1,1), (1,-1,-1) & (0,-1,1), (0,1,-1) \\ \hline
\textbf{Disallowed VEV in NH} & \textbf{Disallowed VEV in IH} \\  \hline
(1,0,0) , (0,1,0), (0,0,1), (1,1,0), & (1,0,0) , (0,1,0), (0,0,1), (1,1,0), \\   
(1,0,1), (1,-1,0), (1,0,-1), (-1,0,1), & (1,0,1), (0,1,1), (1,-1,0), (1,0,-1), \\ 
(-1,1,0), (0,-1,1), (0,1,-1), (-1,1,-1), &  (-1,0,1), (-1,1,0), (1,-1,-1), (-1,1,-1), \\ 
(-1,-1,1), (1,1,-1), (1,-1,1), (-1,-1,0), & (-1,-1,1), (1,1,-1), (1,-1,1), (-1,1,1),   \\ 
(-1,0,-1), (-1,0,0), (0,-1,0), (0,0,-1), & (-1,-1,0), (-1,0,-1), (0,-1,-1), (-1,0,0),  \\ 
(1,1,1), (-1,-1,-1) & (0,-1,0), (0,0,-1), (1,1,1), (-1,-1,-1) \\ \hline

\end{tabular}
\end{small}}
\label{vevalignmentcases}
\end{table}


\begin{table}
\caption{Allowed $ 3\sigma $ range of neutrino oscillation parameters obtained from our ISS model}
{
\begin{adjustbox}{width=15cm}
\begin{tabular}{|c|c|c|c|c|c|}
\hline 
 \multicolumn{6}{|c|}{1. VEV: (0,-1,-1) or (0, 1,1) with normal mass hierarchy} \\ 
\hline 
$ m_{1} $ (eV)  & $ m_{ee} $   &  $Sin\alpha$ & $ Sin \beta $ &   $ Sin\delta_{CP} $ & $  \theta_{23}$    \\ 
\hline
0.01527 $ \rightarrow $ 0.01762 & 0.01599 $ \rightarrow $ 0.01760 & -0.0023 $ \rightarrow $  0.0024 & -1. $ \rightarrow $  0.2449 & -0.2323 $ \rightarrow $  1. & 43.17 $ \rightarrow $  44.07 (lower octant)\\
\hline
 \multicolumn{6}{|c|}{2. VEV: (1,-1,-1) or (-1, 1,1) with normal mass hierarchy} \\ 
 \hline
 $ m_{1} $ (eV)  & $ m_{ee} $   &  $Sin\alpha$ & $ Sin \beta $ &   $ Sin\delta_{CP} $ & $  \theta_{23}$    \\ 
\hline
0.00243 $ \rightarrow $  0.00366 & $8.24651\times 10^{-6}$ $ \rightarrow $  0.00094 & -0.3191 $ \rightarrow $  1. & -0.4782 $ \rightarrow $  1. & -0.4860 $ \rightarrow $  0.4980  & 50.06 $ \rightarrow $  50.70 (higher octant)\\
\hline
 \multicolumn{6}{|c|}{3. VEV: (0,-1,1) or (0, 1,-1) with inverted mass hierarchy} \\ 
 \hline
 $ m_{3} $ (eV)  & $ m_{ee} $   &  $Sin\alpha$ & $ Sin \beta $ &   $ Sin\delta_{CP} $ & $  \theta_{23}$    \\ 
\hline
0.02092 $ \rightarrow $  0.02198 & 0.02215 $ \rightarrow $  0.02337 & 0.9996 $ \rightarrow $  1. & -1.  $ \rightarrow $  1. & -0.6903 $ \rightarrow $  0.7008  & 43.09 $ \rightarrow $  43.30 (lower octant)\\
\hline 
 
\end{tabular} 
\end{adjustbox}
\label{tab:para range}}
\end{table}

\section{Discussion on Results}
\label{sec:level5}
From the recent data of T2K \cite{patrick_dunne_2020_3959558} and NO$\nu$A \cite{alex_himmel_2020_3959581} experiments, it is seen that both the experiments prefer normal hierarchy (NH), without any new physics (for standard oscillation picture). However, a slight tension is seen at the $2 \sigma $ level, where T2K prefers $ \delta_{CP} \sim 3\pi/2 $ for NH  which is excluded by NO$ \nu $A at 90$ \% $ confidence level.  NO$ \nu $A, in general, does not have any strong preference for any particular value of the  CPV phase and has its best fit value around  $ \delta_{CP} \sim \pi $ for NH. This discrepancy is perhaps due to the configuration of baselines and the effect of matter density as neutrinos in NO$ \nu $A experience a much stronger matter effect, and hope that this discrepancy can be alleviated from the robust data obtained by forthcoming neutrino experiments. A slight preference of IH over NH (without new physics) is seen when the recent data of T2K and NO$ \nu $A are combined \cite{Denton:2020uda} , though data from Super Kamiokande still prefer NH over IH \cite{Kelly:2020fkv} . However, it is seen that in the combined experimental data of T2K, NOvA, Super-K \cite{Takeuchi:2020slv} , the NH is preferred over the IH (please see \cite{Denton:2020uda, Kelly:2020fkv, Esteban:2020itz}), and this feature is also observed in our analysis. From  Table (\ref{vevalignmentcases}), we can pinpoint the six allowed VEV alignments of $ \Phi_{s} $ as (0,1,1), (0,-1,-1), (1,-1,-1) and (-1,1,1) which favours NH while only two of them, i.e., (0,-1,1) and (0,1,-1) favours IH. The rest of the cases out of the 26 possible cases in Table (\ref{vevalignmentcases}) are rejected as we do not obtain any event point within the $ 3\sigma $ range of the parameters from current experimental data fitting. The number of allowed cases are very few as we are taking only those high-precision solutions whose accuracy is $ < 10^{-5} $ with $10^6$ random points being generated (for each case). \\
\\
Results in the correlation plots in Figs. (\ref{fig13:LSS_NH1-1-1}- \ref{fig17:LSS_IH0-11}) show that our model can predict the values of  the unknown neutrino parameters, the lightest neutrino mass ($m_1$ or
$m_3$, satisfying the limits on sum of the mass of three light neutrinos from cosmological and tritium beta decay experiments) and Majorana phases (which can vary between $0-2\pi$), which can be tested when measured in future. In these plots, we have chosen only those points for which the known parameters like $\theta_{12}$, $\theta_{13}$, $\theta_{23}$, $\Delta m^2_{12}$, $\Delta m^2_{23}$ and $\delta_{CP}$ lie in their $3\sigma$ range of current global best fit values (Please see Table 1). Next, we discuss these results with reference to Octant-MH degeneracy \cite{Cao:2020ans, Bora:2014zwa} . In the future neutrino experiments, if the mass hierarchy is fixed, say if it is normal hierarchy, then our model will be able to predict  precisely the triplet flavon VEV, i.e., the VEV alignment of $ \Phi_{s} $ is  either (0,1,1) with $\theta_{23}<45^{o}$ or (-1,1,1) with $\theta_{23}>45^{o}$. Similarly, if the mass hierarchy comes out to be inverted in the future experiments, then our model would be able to identify the corresponding VEV alignment of $ \Phi_{s} $ as (0,-1,1) with $ \theta_{23} $ in the lower octant and vice-versa. Thus,  the preferred direction along which flavon field vacuum stabilises can be pinpointed once this entanglement among parameters is fixed, and vice-versa (if the preferred VEV of flavon field is known, it can help us resolve the parameter entanglement).  
The global fit data of neutrino oscillation  \cite{deSalas:2020pgw} used in our analysis is summarised in Table (\ref{tab:3}). In this work, they have considered data from T2K, NO$\nu$A as
LBLs, and have presented global analysis on $0\nu\beta\beta$ decay too. The results from nEXO \cite{nEXO:2021ujk} have also been included for the results of $ 0\nu \beta \beta $ decay in Table (\ref{tab:3}).\\
\\
\begin{table}[h]
\caption{Summary of updated values of oscillation parameters and the preference for octant of $ \theta_{23} $, $ \delta_{CP} $ and mass hierarchy taken from various global experiments \cite{deSalas:2020pgw, nEXO:2021ujk} .} 
{
 \scalebox{0.78}{
\begin{tabular}{|c|c|c|c|}
\hline
 Preference for octant   & Preferred value of  &  Favoured mass    & Results  \\
 of atmospheric angle & CPV phase ( $ \delta_{CP} $ ) & ordering MO & from $ 0\nu \beta \beta $ \\
($ \theta_{23} $)  & &(sign of $ \Delta m^{2}_{31}$)  & Experiments\\
  \hline
 LBLs- two   & NO$ \nu $A - preference 
 & Independent analysis of & $ m_{\beta \beta} \leq 104 -228 $ meV\\
 degenerate solutions &  for $\delta_{CP}=0.8 \pi  $, &   both T2K and  NO$ \nu $A   &     \\
for both the  &  disfavouring  a region  & does not show any  &  by GERDA\\
Octants (LO and HO) &  around best fit of T2K $\delta_{CP}=1.5 \pi  $ &  preference for Mass ordering   &     \\
 \hline
 Combination  & Combination  & All LBL data  &  $m_{\beta \beta} \leq 75 - 350 $ meV \\
of all   &  of LBL+reactor- &  favour IO  &     \\
 Acc. LBL + Reactor & CP conserving value  & -as a  &  by CUORE\\
  & $ \delta_{CP}=0 $ is disfavoured,  & consequence of    &     \\
  - shifts the best fit HO   &  while other CP conserving values  & tension in T2K  & \\
   & $ \delta_{CP}=\pi $ is still allowed  & and NO$ \nu $A data    &     \\
  \hline
 Combination of all   & - & Combination of all   & $m_{\beta \beta} \leq 61 - 165 $ meV \\
  &   &    &     \\
Acc. LBL +atm data (SK)&  & LBL +
reactor data favour NO  &  by KamLand Zen \\
 &   &  Atmospheric SK data favour NO,  &     \\
 shifts preference to HO  &  & whole combination of  & \\
  &   &  LBL+SK favour NO   &     \\
   \hline
  Best fit -  &  Best fit - & Best fit is for NO . & nEXO      \\
  $Sin^{2}\theta_{23} = 0.574(0.578) $  & $\delta_{CP} = 1.08 \pi(1.58 \pi) $ &  A small tension in IO & 1. Sensitivity at 90$\%$ CL\\
   &   &    &     \\
    for NO(IO)  &   for NO(IO) & & $m_{\beta \beta} \leq 4.7 - 26$ meV for NH.\\
    & & & $m_{\beta \beta} < 15$ meV for IH.\\
     &   &    &     \\
       & & & 2. Discovery potential at\\
         & & & ($ 3\sigma $): $m_{\beta \beta} \leq  5.0 - 26$ meV for NH. \\
          &   &    &     \\
         & & &  $m_{\beta \beta} <15$ meV for IH. \\
          &   &    &     \\
\hline
\end{tabular}}
\label{tab:3}}
\end{table}

\begin{figure*}
  \centering
\includegraphics[scale=0.6]{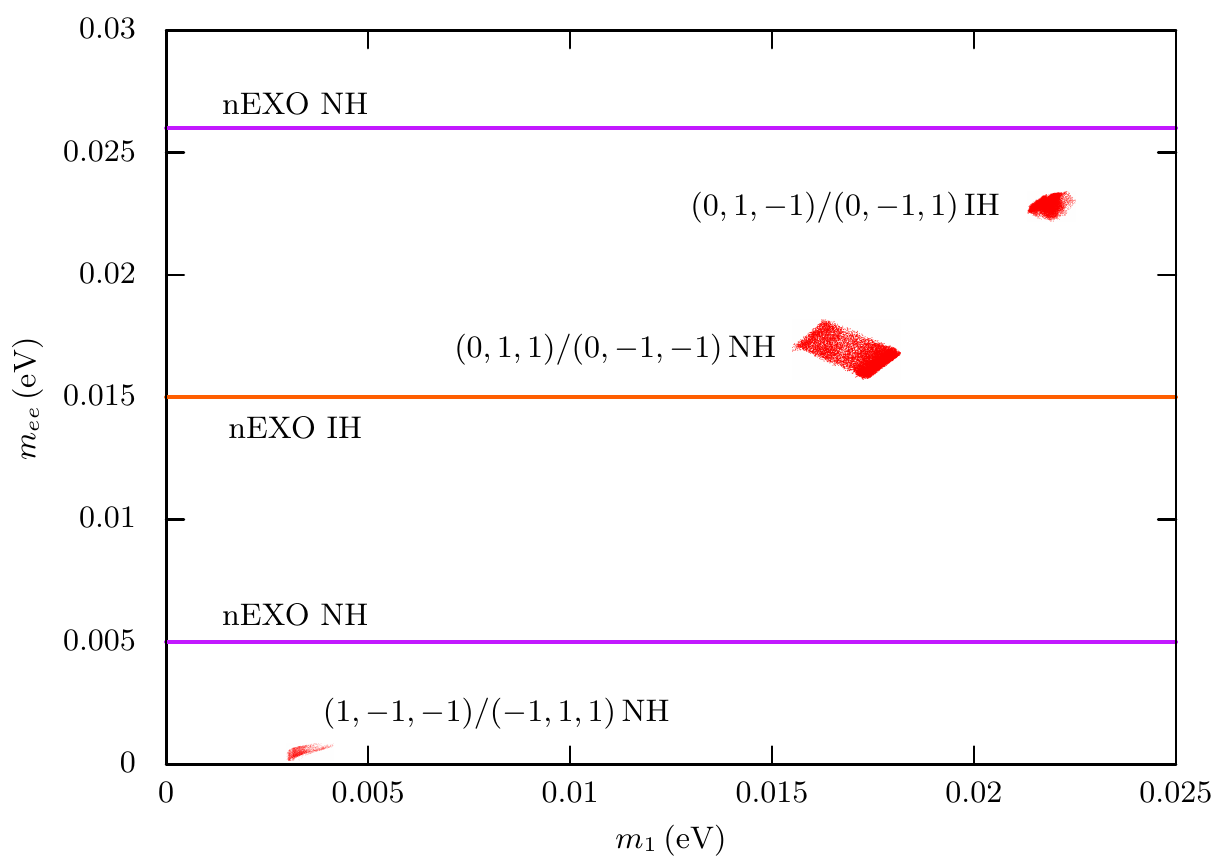}
\caption{ Our predictions on $m_{ee}$ are compared with $3\sigma$ discovery potential limits of nEXO experiment for both NH and IH (see Table 5),  allowed region for NH is below two purple lines, while that for IH is below red line. It can be seen that our results for all the three cases - (0,1,1)NH, (1,-1,-1) NH and (0,1,-1) IH lie in the  $3\sigma$ allowed region of nEXO. We note that the limits of other experiments are more in value than that of nEXO, hence we can say that all the three cases results lie in allowed region of all the experiments listed in Table 5. Also, since the allowed limits of other experiments are much larger than that of nEXO, so they all could not be presented in this same plot due to variation in scale, hence we showed our results with only nEXO limits.}
\end{figure*}
From a careful comparison of our results in Fig(\ref{fig13:LSS_NH1-1-1}-\ref{fig17:LSS_IH0-11}) with Table (\ref{tab:3}), we find that their preferences for NH and HO (higher octant) indicates that the VEV (1,-1,-1)/(-1,1,1) of the $A_4$ triplet flavon field is favoured (case B in Table (\ref{tab:para range})).  Next, we use this novel information  to predict very precise values of $m_{lightest}$ and $m_{ee}$ in our model,  which corresponds to upper right panel (NH, HO) in Fig. (\ref{fig:LSS 0vbb}) - i.e., $m_{lightest} = (0.00243-0.00366)$ eV, and $m_{ee}=(8.24\times 10^{-6}-0.00094)$ eV, which falls in the allowed region as shown in Table 5. Moreover, it is interesting to note  that the preferred mass ordering as well as the  range of lightest neutrino mass and that of $ m_{ee} $ obtained for the allowed cases in our model lie within the range allowed from experiments such as GERDA \cite{GERDA:2019ivs} , CUORE \cite{CUORE:2019yfd} and KamLAND-Zen \cite{KamLAND-Zen:2016pfg} for $ ^{76} $Ge,  for $ ^{130} $Te and for $ ^{136} $Xe respectively at $ 90\% $ confidence level. In Fig. 5, we have compared our results of Fig. 4 with latest $3\sigma$ discovery potential limits from nEXO experiment. Here, allowed region for NH is below two purple lines, while that for IH is below red line (please see Table 5). It can be seen that our results for all the three cases - (0,1,1)NH, (1,-1,-1) NH and (0,1,-1) IH lie in the  $3\sigma$ allowed region of nEXO. We note that the limits of other experiments are more in value than that of nEXO, hence we can say that all the three cases results lie in allowed region of all the experiments listed in Table 5. Hence, our results of Fig (\ref{fig13:LSS_NH1-1-1}-\ref{fig:LSS 0vbb})  show coherence with  reference to that of recent experimentally allowed regions, and so our model is testable with new predictions.\\
\section{Conclusion}
\label{sec:level6}
To conclude, in this work, we presented new ideas on how the resolution of Octant-MH degeneracy present in the measurements of disappearance channel of LBL experiments can be used to get novel information on the VEV alignment of the triplet flavon field of $A_4$ flavour symmetry. We analysed the viability of our ISS model with $  A_{4}\times U(1)_{X} \times Z_{5} \times Z_{4} $ symmetry by determining both the Dirac ($ \delta_{CP} $) as well as the Majorana ($ \alpha $, $ \beta $) phases and the corresponding lightest mass eigenvalue of active neutrinos, keeping the mixing angles and mass squared differences constrained within allowed $3 \sigma $ range. Since computation was done for the accuracy  $<10^{-5}$, we obtained solution points for a narrow range of neutrino oscillation parameters which depicts a very precise solution. The correlation among $ m_{lightest} $ and $ m_{ee} $ of $ 0\nu\beta \beta $ decay agrees very well with their current experimental $3 \sigma $ bounds and sensitivity limits for corresponding MH. By carefully comparing our results with a very recent global analysis (which seems to favour NH and HO), we pinpointed the favoured VEV alignment of $A_4$ triple flavon to be (1,-1,-1)/(-1,1,1), as well as very precise value of $m_{lightest}$ and $m_{ee}$, as shown in Fig. (\ref{fig:LSS 0vbb}) (to be compared with values in Table 5.). Thus the ideas presented here can answer some of the open problems in the neutrino sector and future experiments would be able to confirm or refute the results presented here.

\section*{Acknowledgements}
Authors acknowledge support from FIST and  RUSA grants (Govt. of India) in upgrading the computer laboratory of the department where this work was done.
\appendix
\section{Equations for $A_4$ Flavons present in $ A_{4}\times U(1)_{X} \times Z_{5} \times Z_{4} $ inverse seesaw model}
\label{appen3}
\begin{equation}
\begin{aligned}
\Phi_{a}=\dfrac{J}{3} (-e^{2 i \alpha } m_2 (-\sin (\theta _{23}) \cos (\theta _{12})\\ -e^{i \delta } \sin (\theta _{12}) \sin (\theta _{13}) \cos (\theta _{23})) (\cos (\theta _{12}) \cos (\theta _{23}) \\ -e^{i \delta } \sin (\theta _{12}) \sin (\theta _{13}) \sin (\theta _{23}))+e^{2 i \alpha } m_2 \sin ^2 (\theta _{12}) \cos ^2 (\theta _{13})\\ +m_3 e^{2 i (\beta +\delta )-2 i \delta } \sin ^2(\theta _{13})-m_3 e^{2 i (\beta +\delta )} \sin (\theta _{23}) \cos (\theta _{23}) \cos ^2 (\theta _{13})\\ -m_1 (\sin (\theta _{12}) \sin (\theta _{23})-e^{i \delta } \sin (\theta _{13}) \cos (\theta _{12}) \cos (\theta _{23})) (\sin (\theta _{12}) (-\cos (\theta _{23}))\\ -e^{i \delta } \sin (\theta _{13}) \sin (\theta _{23}) \cos (\theta _{12}))+m_1 \cos ^2 (\theta _{12}) \cos ^2 (\theta _{13}))
\end{aligned}
\label{eq1appen3}
\end{equation}
\begin{equation}
\begin{aligned}
\Phi_{b}=\frac{J}{3} (e^{2 i \alpha } m_2 (\cos (\theta _{12}) \cos (\theta _{23}) -e^{i \delta } \sin (\theta _{12}) \sin (\theta _{13}) \sin (\theta _{23})){}^2 \\ -e^{2 i \alpha } m_2 \sin (\theta _{12}) \cos (\theta _{13}) (-\sin (\theta _{23}) \cos (\theta _{12})-e^{i \delta } \sin (\theta _{12}) \sin (\theta _{13}) \cos (\theta _{23})) \\ +m_3 e^{2 i (\beta +\delta )} \sin ^2 (\theta _{23}) \cos ^2 (\theta _{13})-m_3 e^{2 i (\beta +\delta )-i \delta } \sin (\theta _{13}) \cos (\theta _{13}) \cos (\theta _{23}) \\ +m_1 (\sin (\theta _{12}) (-\cos (\theta _{23}))-e^{i \delta } \sin (\theta _{13}) \sin (\theta _{23}) \cos (\theta _{12})){}^2\\ -m_1 \cos (\theta _{12}) \cos(\theta _{13}) (\sin (\theta _{12}) \sin (\theta _{23})  -e^{i \delta } \sin (\theta _{13}) \cos (\theta _{12}) \cos (\theta _{23})))
\end{aligned}
\label{eq2appen3}
\end{equation}

\begin{equation}
\begin{aligned}
\Phi_{c}=\frac{J}{3} (e^{2 i \alpha } m_2 (-\sin (\theta _{23}) \cos (\theta _{12})-e^{i \delta } \sin (\theta _{12}) \sin (\theta _{13}) \cos (\theta _{23})){}^2 \\ -e^{2 i \alpha } m_2 \sin (\theta _{12}) \cos (\theta _{13}) (\cos (\theta _{12}) \cos (\theta _{23}) -e^{i \delta } \sin (\theta _{12}) \sin (\theta _{13}) \sin (\theta _{23}))\\ +m_3 e^{2 i (\beta +\delta )} \cos ^2 (\theta _{13}) \cos ^2 (\theta _{23}) -m_3 e^{2 i (\beta +\delta )-i \delta } \sin (\theta _{13}) \sin (\theta _{23}) \cos (\theta _{13})\\ +m_1 (\sin (\theta _{12}) \sin (\theta _{23}) -e^{i \delta } \sin (\theta _{13}) \cos (\theta _{12}) \cos (\theta _{23})){}^2 \\ -m_1 \cos (\theta _{12}) \cos (\theta _{13}) (\sin (\theta _{12}) (-\cos (\theta _{23}))  -e^{i \delta } \sin (\theta _{13}) \sin (\theta _{23}) \cos (\theta _{12})))
\end{aligned}
\label{eq3appen3}
\end{equation}

\begin{equation}
\begin{aligned}
\eta=\frac{J}{3}(2 (e^{2 i \alpha } m_2 (-\sin (\theta _{23}) \cos (\theta _{12}) \\ -e^{i \delta } \sin (\theta _{12}) \sin (\theta _{13}) \cos (\theta _{23})) (\cos (\theta _{12}) \cos(\theta _{23}) \\ -e^{i \delta } \sin (\theta _{12}) \sin (\theta _{13}) \sin (\theta _{23}))  +m_3 e^{2 i (\beta +\delta )} \sin (\theta _{23}) \cos (\theta _{23}) \cos ^2 (\theta _{13}) \\ +m_1 (\sin (\theta _{12}) \sin (\theta _{23})-e^{i \delta } \sin (\theta _{13}) \cos (\theta _{12}) \cos (\theta _{23})) (\sin (\theta _{12}) (-\cos (\theta _{23}))\\ -e^{i \delta } \sin (\theta _{13}) \sin (\theta _{23}) \cos (\theta _{12})))+e^{2 i \alpha } m_2 \sin ^2(\theta _{12}) \cos ^2 (\theta _{13}) \\ +m_3 e^{2 i (\beta +\delta )-2 i \delta } \sin ^2(\theta _{13})+m_1 \cos ^2 (\theta _{12}) \cos ^2 (\theta _{13}))
\end{aligned}
\label{eq4appen3}
\end{equation}

\begin{equation}
\begin{aligned}
\rho=\frac{J}{3} (2 (e^{2 i \alpha } m_2 \sin (\theta _{12}) \cos (\theta _{13}) (-\sin (\theta _{23}) \cos (\theta _{12}) \\ -e^{i \delta } \sin (\theta _{12}) \sin (\theta _{13}) \cos (\theta _{23}))+m_3 e^{2 i (\beta +\delta )-i \delta } \sin (\theta _{13}) \cos (\theta _{13}) \cos (\theta _{23}) \\ +m_1 \cos (\theta _{12}) \cos (\theta _{13}) (\sin (\theta _{12}) \sin (\theta _{23}) -e^{i \delta } \sin (\theta _{13}) \cos (\theta _{12}) \cos (\theta _{23}))) \\ +e^{2 i \alpha } m_2 (\cos (\theta _{12}) \cos (\theta _{23})-e^{i \delta } \sin (\theta _{12}) \sin (\theta _{13}) \sin (\theta _{23})){}^2 \\ +m_3 e^{2 i (\beta +\delta )} \sin ^2(\theta _{23}) \cos ^2(\theta _{13})  +m_1 (\sin (\theta _{12}) (-\cos (\theta _{23})) \\ -e^{i \delta } \sin (\theta _{13}) \sin (\theta _{23}) \cos (\theta _{12})){}^2)
\end{aligned}
\label{eq5appen3}
\end{equation}
\begin{equation}
\begin{aligned}
\xi=\frac{J}{3} (2 (e^{2 i \alpha } m_2 \sin (\theta _{12}) \cos (\theta _{13}) (\cos (\theta _{12}) \cos (\theta _{23}) \\ -e^{i \delta } \sin (\theta _{12}) \sin (\theta _{13}) \sin (\theta _{23}))+m_3 e^{2 i (\beta +\delta )-i \delta } \sin (\theta _{13}) \sin (\theta _{23}) \cos (\theta _{13}) \\ +m_1 \cos (\theta _{12}) \cos (\theta _{13}) (\sin (\theta _{12}) (-\cos (\theta _{23}))-e^{i \delta } \sin (\theta _{13}) \sin (\theta _{23}) \cos (\theta _{12})) \\ +e^{2 i \alpha } m_2 (-\sin (\theta _{23}) \cos (\theta _{12})-e^{i \delta } \sin (\theta _{12}) \sin (\theta _{13}) \cos (\theta _{23})){}^2 \\ +m_3 e^{2 i (\beta +\delta )} \cos ^2(\theta _{13}) \cos ^2(\theta _{23})  +m_1 (\sin (\theta _{12}) \sin (\theta _{23}) \\ -e^{i \delta } \sin (\theta _{13}) \cos (\theta _{12}) \cos (\theta _{23})){}^2)
\end{aligned}
\label{eq6appe3n}
\end{equation}
\begin{center}
where, $ J=\dfrac{y^{2}_{m}}{y^{2}_{d}y_{\mu}}[\dfrac{\Lambda^{4}}{v^{2}_{h}v_{\kappa^{\prime}}v^{\dagger}_{\kappa^{\prime \prime}}}] $. We have absorbed $ v_{s} $ in $ \Phi_{a},\Phi_{b},\Phi_{c} $.
\end{center}


\section{Scalar potential minimisation of the model and possible solutions of triplet scalar field, $ \Phi_{s} $}

\label{append 5 : ISS}
We present here the minimisation of the scalar potential considered in our model and later include the all possible solutions of the triplet scalar flavon, $ \Phi_{s} $ after its minimisation:
\begin{eqnarray}
V = V(h) + V(\Phi_{t})+V(\Phi_{s})+V(\eta) \nonumber
  +V(\xi)+V(\rho)+V(\kappa)+ 
V(\kappa^{\prime}) +V(\kappa^{\prime\prime}) \nonumber \\
+V(h, \Phi_{t},\Phi_{s},\eta,\xi,\rho,\kappa,\kappa^{\prime},\kappa^{\prime\prime}) \nonumber 
+V(\Phi_{t},\Phi_{s},\eta,\xi,\rho,\kappa,\kappa^{\prime},\kappa^{\prime\prime})+ \nonumber 
V_{ex}(h,\Phi_{t},\Phi_{s},\eta,\xi,\rho,\kappa,\kappa^{\prime},\kappa^{\prime\prime})
\label{ISS:scalar}
\end{eqnarray}
where, \begin{equation}
V(h)= \mu^{2}_{h} h^{\dagger}h +\lambda_{h}(h^{\dagger}h)(h^{\dagger}h)
\end{equation}

\begin{equation}
\begin{aligned}
V(\Phi_{s})=-\mu^{2}_{s}[\Phi^{\dagger}_{a} \Phi_{a} + \Phi^{\dagger}_{b} \Phi_{c} + \Phi^{\dagger}_{c} \Phi_{b}]  + \lambda_{s}[(\Phi^{\dagger}_{a} \Phi_{a} + \Phi^{\dagger}_{b} \Phi_{c} + \Phi^{\dagger}_{c} \Phi_{b})^{2}  + (\Phi^{\dagger}_{b} \Phi_{b} + \Phi^{\dagger}_{a} \Phi_{c} + \Phi^{\dagger}_{c} \Phi_{a}) (\Phi^{\dagger}_{c} \Phi_{c}\\ + \Phi^{\dagger}_{a} \Phi_{b} + \Phi^{\dagger}_{b} \Phi_{a}) +(2\Phi^{\dagger}_{a} \Phi_{a} - \Phi^{\dagger}_{b} \Phi_{c}  + \Phi^{\dagger}_{c} \Phi_{b})^{2} +2(2 \Phi^{\dagger}_{c} \Phi_{c} - \Phi^{\dagger}_{a} \Phi_{b}  - \Phi^{\dagger}_{b} \Phi_{a})(2 \Phi^{\dagger}_{b} \Phi_{b} - \Phi^{\dagger}_{a} \Phi_{c} - \Phi^{\dagger}_{c} \Phi_{a})
\label{phiS:ISS}
\end{aligned}
\end{equation} 
\\
1. $(\Phi _a,\Phi _b,\Phi _c) \to (0,0,0)  $ \\
2. $(\Phi _a,\Phi _b,\Phi _c) \to ( 0,1,-1)*( -\frac{0.242536 i \mu _s}{\sqrt{\lambda _s}}) $ \\
3.$(\Phi _a,\Phi _b,\Phi _c) \to ( 0,-1,1)*(- \frac{0.242536 i \mu _s}{\sqrt{\lambda _s}}) $\\
4. $ (\Phi _a,\Phi _b,\Phi _c) \to (0,-\frac{(0.210042\, +0.121268 i) \mu _s}{\sqrt{\lambda _s}}, -\frac{(0.210042\, -0.121268 i) \mu _s}{\sqrt{\lambda _s}}) $\\
5. $ (\Phi _a,\Phi _b,\Phi _c) \to (0, \frac{(0.210042\, +0.121268 i) \mu _s}{\sqrt{\lambda _s}}, \frac{(0.210042\, -0.121268 i) \mu _s}{\sqrt{\lambda _s}} )$\\
6. $ (\Phi _a,\Phi _b,\Phi _c) \to (0, \frac{(0.210042\, -0.121268 i) \mu _s}{\sqrt{\lambda _s}}, \frac{(0.210042\, +0.121268 i) \mu _s}{\sqrt{\lambda _s}}) $
\\
7. $ (\Phi _a,\Phi _b,\Phi _c) \to (0, -\frac{(0.210042\, -0.121268 i) \mu _s}{\sqrt{\lambda _s}},  -\frac{(0.210042\,  +0.121268 i) \mu _s}{\sqrt{\lambda _s}})$\\
8. $(\Phi _a,\Phi _b,\Phi _c) \to (1,1,1)*-\frac{0.288675 \mu _s}{\sqrt{\lambda _s}}$\\
9. $ (\Phi _a,\Phi _b,\Phi _c) \to( -\frac{0.288675 \mu _s}{\sqrt{\lambda _s}}, \frac{(0.144338\, +0.25 i) \mu _s}{\sqrt{\lambda _s}},  \frac{(0.144338\, -0.25 i) \mu _s}{\sqrt{\lambda _s}}) $\\
10. $(\Phi _a,\Phi _b,\Phi _c) \to (-\frac{0.288675 \mu _s}{\sqrt{\lambda _s}}, \frac{(0.144338\, -0.25 i) \mu _s}{\sqrt{\lambda _s}},  \frac{(0.144338\, +0.25 i) \mu _s}{\sqrt{\lambda _s}}) $\\
11. $(\Phi _a,\Phi _b,\Phi _c) \to (1,1,1)*\frac{0.288675 \mu _s}{\sqrt{\lambda _s}}$\\
12. $ (\Phi _a,\Phi _b,\Phi _c) \to (\frac{0.288675 \mu _s}{\sqrt{\lambda _s}}, -\frac{(0.144338\, +0.25 i) \mu _s}{\sqrt{\lambda _s}},  -\frac{(0.144338\, -0.25 i) \mu _s}{\sqrt{\lambda _s}}) $\\
13. $ (\phi _a,\phi _b,\phi _c) \to (\frac{0.288675 \mu _s}{\sqrt{\lambda _s}}, -\frac{(0.144338\, -0.25 i) \mu _s}{\sqrt{\lambda _s}},  -\frac{(0.144338\, +0.25 i) \mu _s}{\sqrt{\lambda _s}}) $\\
14. $ (\phi _a,\phi _b,\phi _c) \to (1, 0, 0)*-\frac{0.316228 \mu _s}{\sqrt{\lambda _s}} $\\
15. $ (\phi _a,\phi _b,\phi _c) \to (-1,2,2)* \frac{0.105409 \mu _s}{\sqrt{\lambda _s}} $\\
16. $  (\phi _a,\phi _b,\phi _c)  \to( -\frac{0.105409 \mu _s}{\sqrt{\lambda _s}}, -\frac{(0.105409\, -0.182574 i) \mu _s}{\sqrt{\lambda _s}}, -\frac{(0.105409\, +0.182574 i) \mu _s}{\sqrt{\lambda _s}}) $\\
17. $ (\phi _a,\phi _b,\phi _c) \to( -\frac{0.105409 \mu _s}{\sqrt{\lambda _s}}, -\frac{(0.105409\, +0.182574 i) \mu _s}{\sqrt{\lambda _s}}, -\frac{(0.105409\, -0.182574 i) \mu _s}{\sqrt{\lambda _s}} ) $\\
18. $ (\phi _a,\phi _b,\phi _c)  \to  (1,-2,-2)* \frac{0.105409 \mu _s}{\sqrt{\lambda _s}} $\\
19. $  (\phi _a,\phi _b,\phi _c)  \to (\frac{0.105409 \mu _s}{\sqrt{\lambda _s}}, \frac{(0.105409\, -0.182574 i) \mu _s}{\sqrt{\lambda _s}},  \frac{(0.105409\, +0.182574 i) \mu _s}{\sqrt{\lambda _s}} )$\\
20. $  (\phi _a,\phi _b,\phi _c)  \to \frac{0.105409 \mu _s}{\sqrt{\lambda _s}}, \frac{(0.105409\, +0.182574 i) \mu _s}{\sqrt{\lambda _s}},  \frac{(0.105409\, -0.182574 i) \mu _s}{\sqrt{\lambda _s}} ) $\\
21. $  (\phi _a,\phi _b,\phi _c)  \to (1, 0, 0)*\frac{0.316228 \mu _s}{\sqrt{\lambda _s}} $\\
22. $  (\phi _a,\phi _b,\phi _c) \to (-2,1,1)* \frac{0.140028 \mu _s}{\sqrt{\lambda _s}} $\\
23. $  (\phi _a,\phi _b,\phi _c) \to (-\frac{0.280056 \mu _s}{\sqrt{\lambda _s}}, -\frac{(0.070014\, +0.121268 i) \mu _s}{\sqrt{\lambda _s}},  -\frac{(0.070014\, -0.121268 i) \mu _s}{\sqrt{\lambda _s}}) $\\
24. $  (\phi _a,\phi _b,\phi _c)  \to (-\frac{0.280056 \mu _s}{\sqrt{\lambda _s}}, -\frac{(0.070014\, -0.121268 i) \mu _s}{\sqrt{\lambda _s}},  -\frac{(0.070014\, +0.121268 i) \mu _s}{\sqrt{\lambda _s}}) $\\
25. $  (\phi _a,\phi _b,\phi _c)\to (2,-1,-1)* \frac{0.140028 \mu _s}{\sqrt{\lambda _s}}  $\\
26. $  (\phi _a,\phi _b,\phi _c) \to (\frac{0.280056 \mu _s}{\sqrt{\lambda _s}}, \frac{(0.070014\, +0.121268 i) \mu _s}{\sqrt{\lambda _s}},  \frac{(0.070014\, -0.121268 i) \mu _s}{\sqrt{\lambda _s}}) $\\
27. $  (\phi _a,\phi _b,\phi _c)  \to (\frac{0.280056 \mu _s}{\sqrt{\lambda _s}}, \frac{(0.070014\, -0.121268 i) \mu _s}{\sqrt{\lambda _s}}, \frac{(0.070014\, +0.121268 i) \mu _s}{\sqrt{\lambda _s}}) $.\\

\appendix

\end{document}